# Magnetic-Superconducting phase-diagram of Eu$_{2-x}$Ce$_x$RuSr$_2$Cu$_2$O$_{10-\delta}$

I. Felner, U. Asaf and E. Galstyan

Racah Institute of Physics, The Hebrew University, Jerusalem, 91904, Israel.

Eu$_{2-x}$Ce$_x$RuSr$_2$Cu$_2$O$_{10-\delta}$ (Ru-2122) is the first Cu-O based system in which superconductivity (SC) in the CuO$_2$ planes and *weak*-ferromagnetism (W-FM) in the Ru sub-lattice coexists. The hole doping in the CuO$_2$ planes, is controlled by appropriate variation of the Ce concentration. SC occurs for Ce contents of 0.4-0.8, with the highest T$_C$=35 K for Ce=0.6. The as-prepared non-SC EuCeRuSr$_2$Cu$_2$O$_{10}$ (x=1) sample exhibits magnetic irreversibility below T$_{irr}$=125 K and orders *anti*-ferromagnetically (AFM) at T$_M$ =165 K. The saturation moment at 5 K is M$_{sat}$=0.89 $\mu_B$ /Ru close to the expected 1 $\mu_B$ for the low-spin state of Ru$^{5+}$. Annealing under oxygen pressures, does not affect these parameters, whereas depletion of oxygen shifts both T$_{irr}$ and T$_M$ up to 169 and 215 K respectively. Systematic magnetic studies on Eu$_{2-x}$Ce$_x$RuSr$_2$Cu$_2$O$_{10-\delta}$ show that T$_M$, T$_{irr}$ and M$_{sat}$ decrease with x, and the Ce dependent magnetic-SC phase diagram is presented. A simple model for the SC state is proposed. We interpret the magnetic behavior in the framework of our ac and dc magnetic studies, and argue that: (i) the system becomes AFM ordered at T$_M$; (b) at T$_{irr}$ < T$_M$, W-FM is induced by the canting of the Ru moments, and (c), at lower temperatures the appropriate samples become SC at T$_C$. The magnetic features are not affected by the SC state, and the two states coexist.

## Introduction

Much attention has been focused on a phase resembling the superconducting $RBa_2Cu_3O_7$ (R=rare-earth) materials, having the composition $R_{2-x}Ce_xMSr_2Cu_2O_{10}$ (M-2122, **M**= Nb, Ru or Ta) [1]. The tetragonal M-2122 structure (space group I4/mmm) evolves from the $RBa_2Cu_3O_7$ structure by inserting a fluorite type $R_{1.5}Ce_{0.5}O_2$ layer instead of the R layer in $RBa_2Cu_3O_7$, thus shifting alternate perovskite blocks by (a+b)/2 . The M ions reside in the Cu (1) site and only one distinct Cu site (corresponding to Cu (2) in $RBa_2Cu_3O_7$) with fivefold pyramidal coordination, exists. The hole doping of the Cu-O planes, which results in metallic behavior and SC, can be optimized with appropriate variation of the R/Ce ratio [2]. SC occurs for Ce contents of 0.4-0.8, and the highest $T_C$ was obtained for Ce=0.6. The Nb-2122 and Ta-2122 materials are SC with $T_C$ 28-30 K.

Coexistence of weak-ferromagnetism (W-FM) and superconductivity (SC) was discovered few years ago in $R_{1.5}Ce_{0.5}RuSr_2Cu_2O_{10}$ (R=Eu and Gd, Ru-2122) layered cuprate systems [3-6], and more recently [7] in $GdSr_2RuCu_2O_8$, (Ru-1212). The SC charge carriers originate from the $CuO_2$ planes and the W-FM state is confined to the Ru layers. In both systems, the magnetic order does not vanish when SC sets in at $T_C$, and remains unchanged and coexists with the SC state. The Ru-2122 materials (for R=Eu) display a magnetic transition at $T_M$= 125-180 K and bulk SC below $T_C$ = 32-50 K ($T_M >T_c$) depending on oxygen concentration and sample preparation [6]. SC survives because the Ru moments probably align in the basal planes, which are practically decoupled from the $CuO_2$ planes, so that there is no pair breaking. Specific heat studies show a sizeable typical jump at $T_C$ and the magnitude of the $\Delta C/T$ (0.08 mJ/gK$^2$) indicates clearly the presence of bulk SC [8]. The specific heat anomaly is independent of the applied magnetic field. Scanning tunneling spectroscopy [3], muon-spin rotation [9] and magneto-optic experiments [10] have demonstrated that all materials are microscopically uniform with no evidence for spatial phase separation of superconducting and magnetic regions. That is, both states coexist intrinsically on the microscopic scale.

In the Ru-2122 system, the W-FM state, as well as irreversibility phenomena, arise as a result of an anti-symmetric exchange coupling of the Dzyaloshinsky-Moriya (DM) type [3] between neighboring Ru moments, induced by a local distortion that breaks the tetragonal symmetry of the $RuO_6$ octahedra. Due to this DM interaction, the field causes the adjacent spins to cant slightly out of their original direction and to align a component of the moments with the direction of the applied field. Below the irreversible temperature

($T_{irr}$), which is defined as the merging temperature of the zero-field (ZFC) and field-cooled (FC) curves, the Ru-Ru interactions begin to dominate, leading to reorientation of the Ru moments, which leads to a peak in the magnetization curves. The Ce concentration affects the hole carrier concentration as a result the SC properties and the $T_C$ values of $R_{2-x}Ce_x\mathbf{Ru}Sr_2Cu_2O_{10-\delta}$ system. The remaining unresolved question concerns the effect of Ce concentration on magnetic properties of this system. It was shown that the peak position of the ZFC magnetization data increases with increasing Ce[11], however, the detailed magnetic features of this system are lacking.

In attempting to understand the mechanism of SC and W-FM in the Ru-2122 system, an approach involving a systematic magnetic properties of $Eu_{2-x}Ce_x\mathbf{Ru}Sr_2Cu_2O_{10-\delta}$ for $0.4<x<1$ was employed. $Eu^{3+}$ is used in order to diminish the paramagnetic contribution of $Gd^{3+}$. This paper is organized as follows: (a) We first show the effect of Ce on the SC properties and compare between the Ru-2122 and $La_{2-x}Sr_xCuO_4$ systems. (b) we present the magnetic properties of the non-SC $EuCe\mathbf{Ru}Sr_2Cu_2O_{10}$ (x=1) material, in which the oxygen concentration is fixed. Annealing under high oxygen pressure does not alter the magnetic properties. (c) We exhibit a systematic study of the Ce concentration, on the magnetic parameters of $Eu_{2-x}Ce_x\mathbf{Ru}Sr_2Cu_2O_{10-\delta}$ and construct the full SC-magnetic phase diagram. (d) The qualitative model of the magnetic structure in the Ru-2122 system is discussed.

## Experimental details

Ceramic samples with nominal composition $Eu_{2-x}Ce_xRuSr_2Cu_2O_{10-\delta}$ (x = 0.4-1) were prepared by a solid-state reaction technique. Prescribed amounts of $Eu_2O_3$, $CeO_2$, $SrCO_3$, Ru and CuO were mixed and pressed into pellets and preheated at 1000° C for about 1 day at atmospheric pressure. The product was cooled, reground and sintered at 1050°C for 50 h in a slightly pressurized oxygen atmosphere (~1.1 atm) and then furnace cooled to ambient temperature (asp samples). All the asp materials reported were prepared at the same time and under the same conditions. Part of the asp $EuCeRuSr_2Cu_2O_{10}$ sample was re-heated for 24 h at 800° C under high oxygen pressure (60 atm.) and another part was quenched from 1050° C to room temperature, denoted as hop and quenched materials, respectively. Determination of the absolute oxygen content in the asp Ru-2122 material, as well as in the hop and quenched samples, is difficult because $CeO_2$ is not completely reducible to a stoichiometric oxide when heated to high temperatures.

Powder X-ray diffraction (XRD) measurements confirmed the purity of the compounds (~97%) and indicate within the instrumental accuracy, that all samples have the same lattice parameters a=3.846(1) Å and c=28.72(1) Å, in excellent agreement with Ref. 3. ZFC and FC dc magnetic measurements in the range of 5-300 K were performed in a commercial (Quantum Design) super-conducting quantum interference device (SQUID) magnetometer. The resistance was measured by a standard four contact probe and the ac susceptibility was measured by a home-made probe with excitation frequency and amplitude of 733 Hz and 30 mOe respectively, both inserted in the SQUID magnetometer.

## Experimental results and discussion

Due to the similarity of the ionic radii of $Eu^{3+}$ (0.94 Å) and $Ce^{4+}$ (0.87 Å) and within the instrumental accuracy, the lattice parameters of $Eu_{2-x}Ce_xRuSr_2Cu_2O_{10-\delta}$, are independent of Ce content. The detailed crystal structure and the atomic positions Ru-2122 were studied by synchrotron X-rays diffraction [12] and neutron diffraction [13] experiments, which show that the $RuO_6$ octahedra are rotated ~ 14° around the c-axis and that this rotation is essentially the same for x=1 and x=0.6 as well as for Ru-1212. There is no evidence for super-cell peaks in the Ru-2122 samples.

### (a) Superconductivity in the $Eu_{2-x}Ce_xRuSr_2Cu_2O_{10-\delta}$ system

The temperature dependence of the normalized ac susceptibility curves (at $H_{dc}$=0) for the $Eu_{2-x}Ce_xRuSr_2Cu_2O_{10-\delta}$ system are presented in Fig. 1. It is readily observed that the x=1 and 0.9 are not SC and that SC occurs for Ce contents of x=0.8-0.4. Fig 2. exhibits the onset of SC deduced from these ac plots which exhibit a bell shape behavior with a peak at 35 K for Ce=0.6. Similar values are obtained by our resistivity measurements (not shown). A similar bell shape behavior was observed in the $Eu_{2-x}Ce_xNbSr_2Cu_2O_{10-\delta}$ (x=0.4-1) system, which serve as reference materials. It is apparent in Fig. 1 that the SC transitions, are much broader than those observed in many other HTSC materials. This transition width is comparable to our previous resistivity data [3,6] and with Ref. 11. Such a broadening is typical of under-doped Cu based high $T_c$ materials where inhomogeneity in the oxygen concentration causes a distribution in the $T_c$ values. In addition, the SC transition width may be due to the spontaneous vortex state discussed in detail in Ref. 4. In the next section we suggest an intuitive explanation as to why the M-2122 (M= Ru, Ta and Nb) materials are superconducting.

Given the variety of crystal structures and the chemical methods used to introduce holes into the $CuO_2$ layers, it is well established that a "generic' electronic phase diagram can be sketched for all compounds. The hole (or carrier) density (p) in the $CuO_2$ planes, or deviation of the formal Cu valence from $Cu^{2+}$, is a primary parameter, which affects $T_C$ in most of the HTSC compounds. In the well-established phase diagram $La_{2-x}Sr_xCuO_4$, the stoichiometric parent $La_2CuO_4$ is AFM and insulating. The magnetic interactions are well described by a simple Heisenberg model, with a large exchange interaction (J= 1500 K) value. In $La_{2-x}Sr_xCuO_4$ the charge carrier concentration, can be varied by replacing $Sr^{2+}$ for $La^{3+}$ (or by removal or addition of oxygen in $YBa_2Cu_3O_z$). The variation of $T_C$ as a function of hole doping exhibits a bell shape behavior, with a peak for the optimally doped material (x=0.15 for $La_{2-x}Sr_xCuO_4$).

The Eu, Ce and Ru ion valencies in $Eu_{1.5}Ce_{0.5}RuSr_2Cu_2O_{10-\delta}$ have been studied by Mossbauer spectroscopy (MS) and X-ray-absorption spectroscopy (XAS) techniques [14-15]. MS performed at 90 and 300 K on $^{151}Eu$ show a single narrow line with an isomer shift of 0.69(2) mm/s, indicating that the Eu ions are trivalent with a nonmagnetic J=0 ground state. XAS taken at $L_{III}$ edges of Ce shows also that Ce is tetravalent. XAS taken at the K edge of Ru, at room temperature indicates clearly that the Ru ions are pentavalent [14-15]. It is also apparent, that bulk SC in the M-2122 system appears only for pentavalent M ions and both Nb-2122 and Ta-2122 materials are SC with $T_C$ 28-30 K [1].

We argue that the $RCeMSr_2Cu_2O_{10}$ (R=Eu or Gd and R/Ce=1) samples serve as the parent stoichiometric insulator compounds (similar to $La_2CuO_4$). Since the valence of $R^{3+}$, $Ce^{4+}$, $Ru^{5+}$ $Sr^{2+}$ $Cu^{2+}$ and $O^{2-}$ are conclusive, a straightforward valence counting yields a fixed oxygen concentration of 10. Hole doping of the Cu-O planes, which results in metallic behavior and SC, can be optimized with appropriate variation of the $R^{3+}$/ $Ce^{4+}$ ratio (trivalent $R^{3+}$ ions are replaced for $Ce^{4+}$). SC occurs for Ce contents of 0.4-0.8, and the optimally doped sample is obtained for Ce=0.6. In fact, unlike the $La_{2-x}Sr_xCuO_4$ system, this substitution does not appear to significantly alter the hole concentration (p) on the Cu-O planes. This is apparent in Fig. 2 which shows that the change of x from 0.8 to 0.6, results in a small increase in $T_C$. Indeed, if all the carriers were induced into the Cu-O planes, then for the under doped (x=0.8) to the optimally doped (x=0.6) samples, p should vary by 0.2 and result in a large shift in $T_C$, as observed in $La_{2-x}Sr_xCuO_4$ and in other HTSC materials. It is thus possible, that in all

M-2122 compounds, the extra holes introduced by reducing the Ce content, are partially compensated for by depletion of oxygen[11] and in $R_{2-x}Ce_xMSr_2Cu_2O_{10-\delta}$ the oxygen deficiency ($\delta$) increases with $R^{3+}$. In the case of M=Ru, EuCe**Ru**Sr$_2$Cu$_2$O$_{10}$ is magnetically ordered at $T_M$ =165 K and the next section deals with its detailed magnetic behavior.

**(b) The effect of oxygen on the magnetic behavior of EuCe*Ru*Sr$_2$Cu$_2$O$_{10}$**

EuCeRuSr$_2$Cu$_2$O$_{10}$ is not SC and is only magnetically ordered (Fig.1). ZFC and FC dc magnetic measurements for the parent as-prepared (asp) sample, were performed over a broad range of applied magnetic fields and typical M/H curves measured at 50 Oe, are shown in Fig.3. The two curves merge at $T_{irr}$=125 K. Note the ferromagnetic-like shape of the FC branches. $T_M(Ru)$ is not at $T_{irr}$. The M/H(T) curves do not lend themselves to an easy determination of $T_M(Ru)$, and $T_M(Ru)$=165 K, was obtained directly from the temperature dependence of the saturation moment ($M_{sat}$), discussed below. [An alternative way to determine $T_M(Ru)$ is to cool the material from above $T_M$, under a small *negative* magnetic field (say -5 Oe), which aligns the Ru sublattice. At low temperatures, a small *positive* (5 Oe) is applied. Due to the high anisotropy, the Ru moments remain opposite the field direction up to $T_M$. The measured negative M(T) curve becomes zero at $T_M$]. $T_{irr}$ is field dependent, and shifted to lower temperatures with the applied field. $T_{irr}$ =91, 64, 39, 28 and 14 K for H= 250, 500, 1000, 2000 and 3000 Oe respectively. For higher external fields the irreversibility is washed out, and both ZFC and FC curves collapse to a single ferromagnetic-like behavior. It appears that the magnetic properties of Ru in EuCeRuSr$_2$Cu$_2$O$_{10}$ are all *enhanced,* as compared to the asp $Eu_{2-x}Ce_xRuSr_2Cu_2O_{10-\delta}$ samples with x<1. The effect of Ce concentration on the magnetic parameters of the Ru sublattice will be discussed below.

Similar M/H(T) curves were performed to the oxygen annealed (hop) sample, and Fig. 3 shows that both $T_{irr}$ and $T_M(Ru)$ remain unchanged. This is in contrast to $Eu_{1.5}Ce_{0.5}RuSr_2Cu_2O_{10-\delta}$ where annealing under the same oxygen pressure affects $T_{irr}$ and $T_M(Ru)$ and shift them to higher temperatures[6]. As stated above, the oxygen concentration in EuCeRuSr$_2$Cu$_2$O$_{10}$ is fixed and does not change during the annealing process. On the other hand, during the quenching process, a small amount of oxygen is depleted and both $T_{irr}$ and $T_M(Ru)$ are shifted to 169 and 215 K respectively (Fig. 3). It reminiscent of the magnetic phase-diagram of YBa$_2$Cu$_3$O$_z$ (z<6.5), where the depletion of oxygen increases the magnetic transition of the CuO$_2$ planes.

M(H) measurements at various temperatures for the asp, hop and quenched samples have been carried out, and the results obtained for the asp sample are exhibited in Figs. 4-6. All M(H) curves below $T_M$, are strongly dependent on the field (up to 2-4 kOe), until a common slope is reached (Fig. 4 inset). M(H) can be described as: $M(H) = M_{sat} + \chi H$, where $M_{sat}$ corresponds to the W-FM contribution of the Ru sub lattice, and $\chi H$ is the linear paramagnetic contribution of Eu and Cu). The saturation moment obtained at 5 K is $M_{sat} = 0.89(1)\mu_B$. Similar M(H) curves have been measured at various temperatures and Fig. 5 (inset) shows the decrease of $M_{sat}$ with temperature. $M_{sat}$ becomes zero at $T_M(Ru)=165(2)$. Similar $M_{sat}$ and $T_M(Ru)$ values were obtained for the hop material. However, for the quenched material, the saturation moment at 5 K remains unchanged, but $T_M$ is shifted to 215 (2) K. Thus, only the magnetic transitions are sensitive to the oxygen concentration. The measured $M_{sat}= 0.89\mu_B$, is somewhat smaller than the fully saturated moment $1\mu_B$ expected for the low-spin state of $Ru^{5+}$, i.e. $g\mu_B S$ for $g=2$ and $S=0.5$. This means that a small canting on adjacent Ru spins occurs, and the saturation moments are not the full moments of the $Ru^{5+}$ ions. The exact nature of the local structural distortions causing DM exchange coupling in Ru-2122 (see above) is not presently known.

At low applied fields, the M(H) curve exhibits a typical ferromagnetic-like hysteresis loop (Fig. 6) similar to that reported in Ref. 3. The positive virgin curve at low fields, indicates clearly that SC is totally suppressed, Two other characteristic parameters of the hysteresis loops are shown in Fig. 6, namely, the remanent moment, ($M_{rem} = 0.41\mu_B/Ru$) and the coercive field ($H_C = -190$ Oe at 5 K). This $M_{rem}$ is much larger than $M_{rem} = 0.035 \mu_B$ obtained for $EuSr_2RuCu_2O_8$ [12]. The same $M_{rem}$ and $H_C$ values (at 5 K) were obtained for the hop and quenched materials. Fig. 4 shows the temperature dependence of $M_{rem}$ (5 K) which disappears at $T_{irr}$. The large $M_{rem}$ (0.41$\mu_B$ at 5 K) in relation to the saturation moment ($M_{sat}= 0.89\mu_B$) is consistent with ferromagnetic-like order in $EuCeRuSr_2Cu_2O_{10}$, although neutron diffraction measurements are required to precisely determine the nature of the magnetic order. M(H) curves measured at various temperatures yield the $M_{rem}(T)$ and $H_C(T)$ values which are plotted in Fig. 5 (inset). For both the asp and quenched samples $M_{rem}(T)$ also disappear at $T_{irr}$, and $H_C(T)$ become zero around 80 K and 130 K respectively (Fig. 5).

Above $T_M(Ru)$, the $\chi(T)$ curve at 10 kOe, for asp $EuCeRuSr_2Cu_2O_{10}$ has the typical paramagnetic shape and adheres closely to the Curie-Weiss (CW) law: $\chi = \chi_0 + C/(T-\theta)$, where $\chi_0$ is the temperature independent part of $\chi$, C is the Curie constant, and $\theta$ is the CW temperature. The net paramagnetic Ru contribution to (T), was obtained by subtracting (T) of EuCe**Nb**Sr$_2$Cu$_2$O$_{10}$ (the reference material) from the measured data. This procedure yields: $\chi_0$ =0.0063 and C=0.57(1) emu/mol Oe and $\theta$= 146(1) K, which corresponds to an effective moment P$_{eff}$ =2.13 $\mu_B$. $\theta$ obtained, is in fair agreement with T$_M$, but P$_{eff}$ is greater than the expected value of the low-spin state of Ru$^{5+}$ and S=0.5 (P$_{eff}$ =1.73 $\mu_B$)..

**(c) The effect of Ce on the magnetic behavior of Eu$_{2-x}$Ce$_x$RuSr$_2$Cu$_2$O$_{10-\delta}$**

As stated above, the compounds described here have been prepared simultaneously under the same conditions. Since the magnetic properties of the Ru-2122 system depend strongly on oxygen concentration[6], the data presented here differ slightly from those reported in our previous publications. ZFC and FC magnetic as well as isothermal M(H) measurements, have been performed on all Eu$_{2-x}$Ce$_x$RuSr$_2$Cu$_2$O$_{10-\delta}$ (x=0.4-1) compounds. Generally speaking, the magnetic behavior of all materials is quite similar to these described in Figs 3-6 and for the sake of brevity, we display the data obtained for x=0.1 and x=0.5 (Fig. 7). It is readily observed, that for x=1, the magnetic properties due to the Ru are all *enhanced*, as compared to the x=0.5. The latter sample is SC and its ZFC branch starts from negative moments. The inflection in the FC branch agrees well with T$_c$ determined from the ac curve (Fig.1). The absence of a complete Meissner effect in the Ru-2122 system is discussed in length elsewhere. However this is of little interest in the present discussion.

The variation of T$_{irr}$ and T$_M$ as a function of Ce concentration in Eu$_{2-x}$Ce$_x$**Ru**Sr$_2$Cu$_2$O$_{10-\delta}$, is summarized in Fig. 8. The enhancement of the magnetic properties is manifested by the monotonic rise of T$_{irr}$ and T$_M$ as x increases. M$_{sat}$ values at 5 K, increase gradually with x, M$_{sat}$ =0.43, 0.46, 0.60 0.67 and 0.86$\mu_B$ for x=0.4, 0.6, 0.7 0.8 and 0.9 respectively. It is apparent that this trend is not affected by the SC state which is induced for x=0.8. Due to the presence of a tiny amount of SrRuO$_3$ in the x=0.4 sample (not detectable by XRD), its T$_M$ was not determined.

In the paramagnetic range, the parameters extracted using the CW law, have been obtained by subtracting the paramagnetic moment of the relevant Nb-2122 compound, as described above. It appears that the C and $\theta$ parameters for all materials, are very close to those of EuCe**Ru**Sr$_2$Cu$_2$O$_{10}$ ; i.e. for x=0.5 C=0.58 emu/mol Oe, (P$_{eff}$=2.15 $\mu_B$) and $\theta$= 134(1) K), indicating similar net paramagnetic Ru contribution in all Ru-2122 compounds.

**(d) The qualitative magnetic structure of $Eu_{2-x}Ce_xRuSr_2Cu_2O_{10-\delta}$**

All magnetic parameters (such as $T_M$, $T_{irr}$ and $M_{sat}$) are shifted to higher values with increasing Ce. Our general picture is that in $Eu_{2-x}Ce_x\mathbf{Ru}Sr_2Cu_2O_{10-\delta}$ all compounds have a similar magnetic structure. We assume that the small difference between $Eu^{3+}$ and $Ce^{4+}$ ionic radii discussed above, decreases the mean Ru-Ru distance, and as a result the magnetic exchange interactions become stronger with Ce.

While our data described here do not include any determination of the magnetic structural order of the Ru sublattice in Ru-2122, the results are compatible with a simple model which is, however, of use in understanding the qualitative features at low applied fields. Starting from high to low temperatures, the magnetic behavior is basically divided into 4 regions. (i) At elevated temperatures, the paramagnetic net Ru moment is well described by the CW law, and the extracted $P_{eff}$=2.15 $\mu_B$ and $\theta$=134-146 K values, practically do not alter with Ce. (ii) At $T_M$ (depends on Ce content Fig. 8) which is deduced directly from the temperature dependence $M_{sat}$ (Fig. 5), the Ru sub lattice becomes AFM ordered. (iii) At $T_{irr}$ < $T_M$, which is Ce dependent (Fig. 8) and also varies with the external field, a weak ferromagnetism is induced, which originates from canting of the Ru moments. $T_{irr}$ is defined as the merging point of the low field ZFC and FC branches, or alternatively, at the temperature in which the remanent moment disappears. This canting arises from the DM anti-symmetric super-exchange interaction, which by symmetry, follows from the fact that the $RuO_6$ octahedra tilt away from the crystallographic c axis [3,11]. At high magnetic field (H>3000 Oe) the irreversibility is washed out and the M(T) curves exhibit a ferromagnetic-like behavior. (iv) At lower temperatures SC is induced. Fig. 2 shows that $T_C$ depends strongly on the $R^{3+}/Ce^{4+}$ (as hole carriers) and on oxygen concentrations [6]. Below $T_C$, both SC and weak-ferromagnetic states coexist [3,10] and the two states are practically decoupled. This model is supported by our unpublished non-linear ac susceptibility measurements, which show non-linear signals up to $T_M$, and also from Mossbauer studies on $^{57}Fe$ doped material[17]. However, the present interpretation, differs completely from the phase separation of AFM and FM nano-domain particle scenario, suggested in Ref 16. Neutron diffraction measurements are required to precisely determine the nature of the magnetic order in the Ru-2122 system.

In conclusion, the magnetic insulator parent $EuCe\mathbf{Ru}Sr_2Cu_2O_{10}$ (x=1), is used to describe the coexistence of both SC and weak-ferromagnetism states in $Eu_{2-x}Ce_x\mathbf{Ru}Sr_2Cu_2O_{10-\delta}$ (x=0.4-1). For x=1 ($\delta$=0), annealing under high oxygen pressure does not affect the magnetic properties, whereas $T_M$ is enhanced by oxygen depletion. Hole doping of the Cu-O planes, which results in SC, can be optimized with appropriate variation of the $Eu^{3+}/Ce^{4+}$ ratio and the optimally doped material is obtained for Ce=0.6. The magnetic structure of all materials studied is practically the same, but the magnetic parameters, such as $T_M$ and $M_{sat}$, decrease with decreasing Ce content. Two steps in the magnetic behavior are presented. At $T_M$ ranging from 125 K (for x=0.5) to165 K (for x=1) all materials become AFM ordered. At $T_{irr}$ (depends on Ce) a W-FM state is induced, originating from canting of Ru moments. This canting arises from the DM anti-symmetric super-exchange interaction and follows from the fact that the $RuO_6$ octahedra tilt away from the crystallographic c axis. A direct magnetic structure determination by neutron diffraction or $^{99}$Ru Mossbauer spectroscopy studies are warranted to confirm our assumptions.

**Acknowledgments** This research was supported by the Israel Academy of Science and Technology and by the Klachky Foundation for Superconductivity.

**Figures Captions**

Fig. 1. Normalized ac susceptibility, measured at $H_{dc}=0$ of the asp $Eu_{2-x}Ce_xRuSr_2Cu_2O_{10-}$ samples. Note, that the two x=1 and x=0.9 materials are magnetically ordered only.

Fig.2. The bell shape SC onset temperature as a function of Ce content.

Fig. 3. ZFC and FC susceptibility curves for $EuCeRuSr_2Cu_2O_{10}$ samples, asp, annealed under 60 oxygen atmosphere and quenched materials measured at 50 Oe.

Fig. 4. The temperature dependence of 5 K remanent moment of $EuCeRuSr_2Cu_2O_{10}$. The inset shows the high field magnetization and the saturation moment at 5 K.

Fig. 5. The coercive field $H_C$ as function of temperature for asp and quenched $EuCeRuSr_2Cu_2O_{10}$ samples. The inset shows the temperature dependence of the saturation and the remanent moments.

Fig. 6. The hysteresis low field loop at 5 K.

Fig. 7. ZFC and FC susceptibility curves for asp x=1 and x=0.5 samples, measured at 50 Oe.

Fig. 8. $T_M$ and $T_{irr}$ as function of Ce in $Eu_{2-x}Ce_xRuSr_2Cu_2O_{10}$.

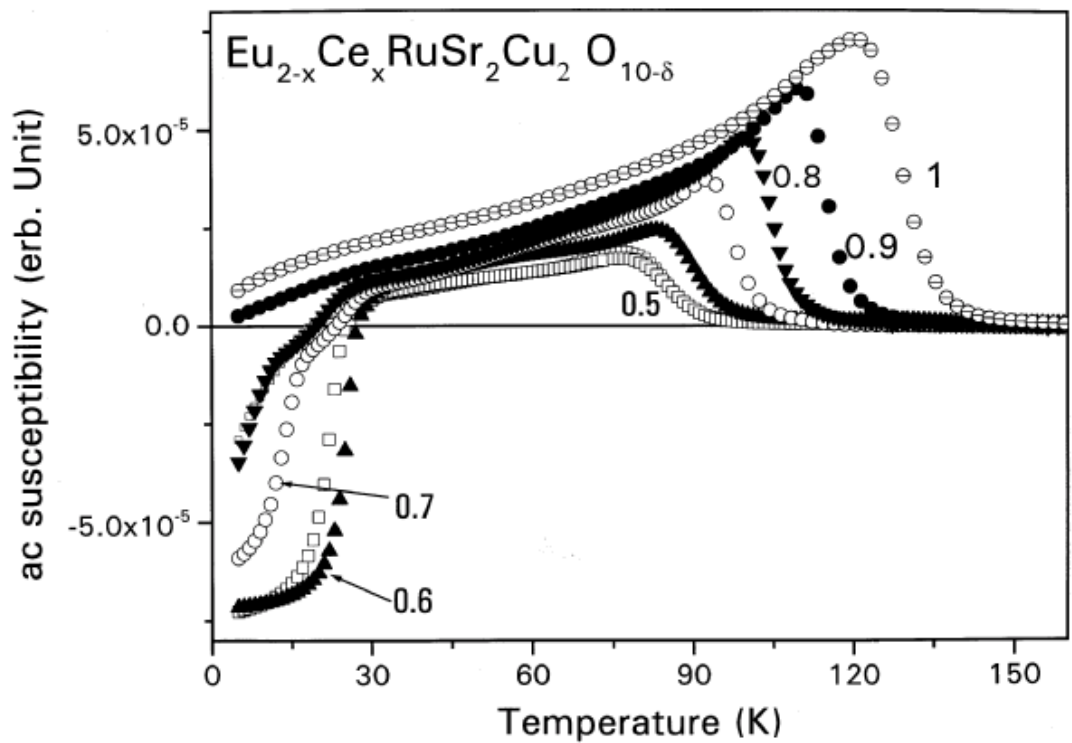

**Fig. 1**. Normalized ac susceptibility, measured at $H_{dc}=0$ of the asp $Eu_{2-x}Ce_xRuSr_2Cu_2O_{10-\delta}$ samples. Note, that the two x=1 and x=0.9 materials are magnetically ordered only

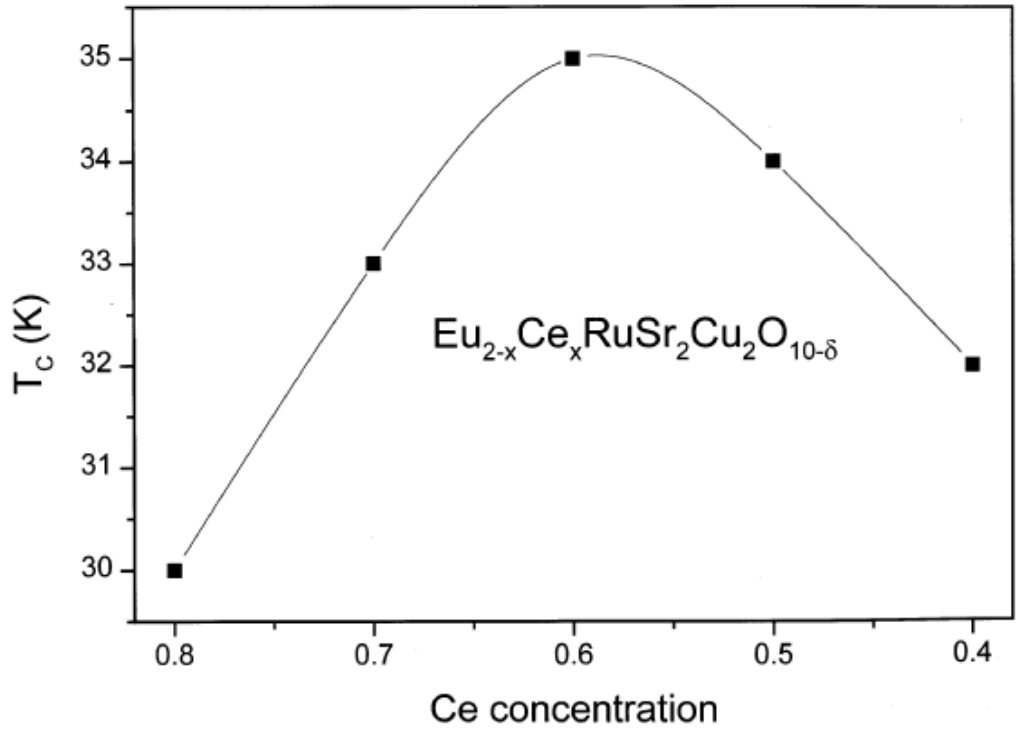

Fig.2. The bell shape SC onset temperature as a function of Ce content.

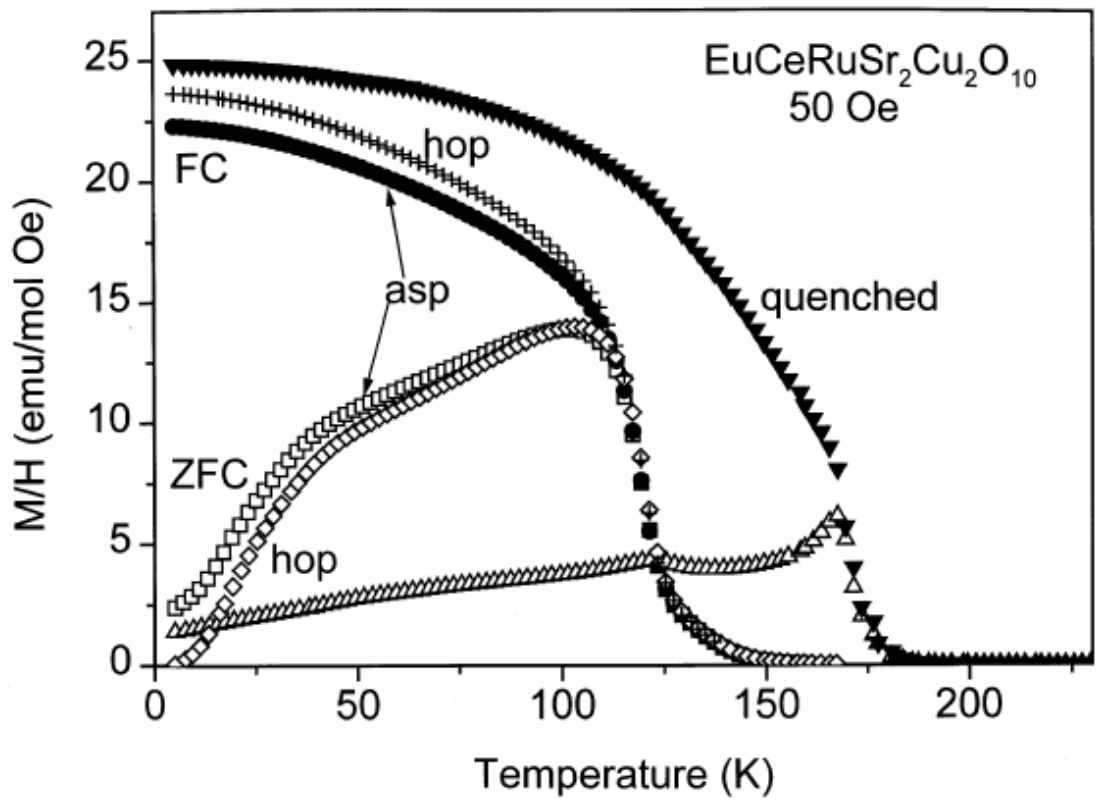

Fig. 3. ZFC and FC susceptibility curves for EuCeRuSr$_2$Cu$_2$O$_{10}$ samples, asp, annealed under 60 oxygen atmosphere (hop) and quenched materials measured at 50 Oe.

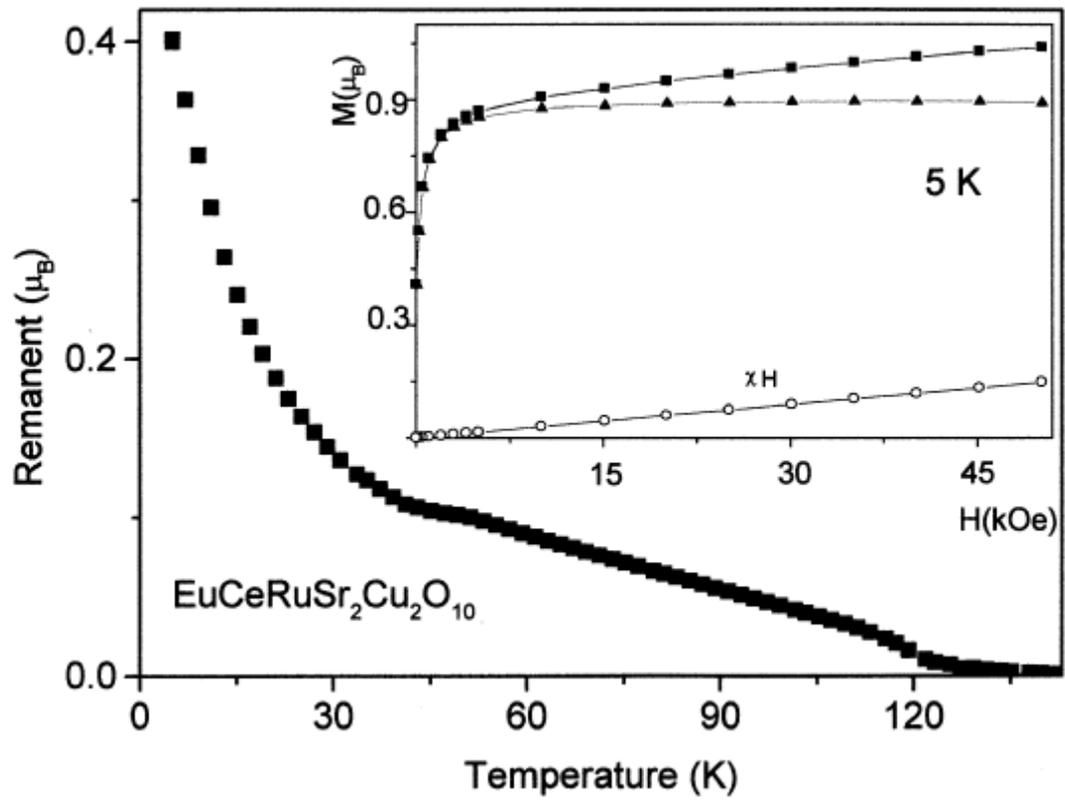

Fig. 4. The temperature dependence of 5 K remanent moment of EuCeRuSr$_2$Cu$_2$O$_{10}$. The inset shows the high field magnetization and the saturation moment at 5 K.

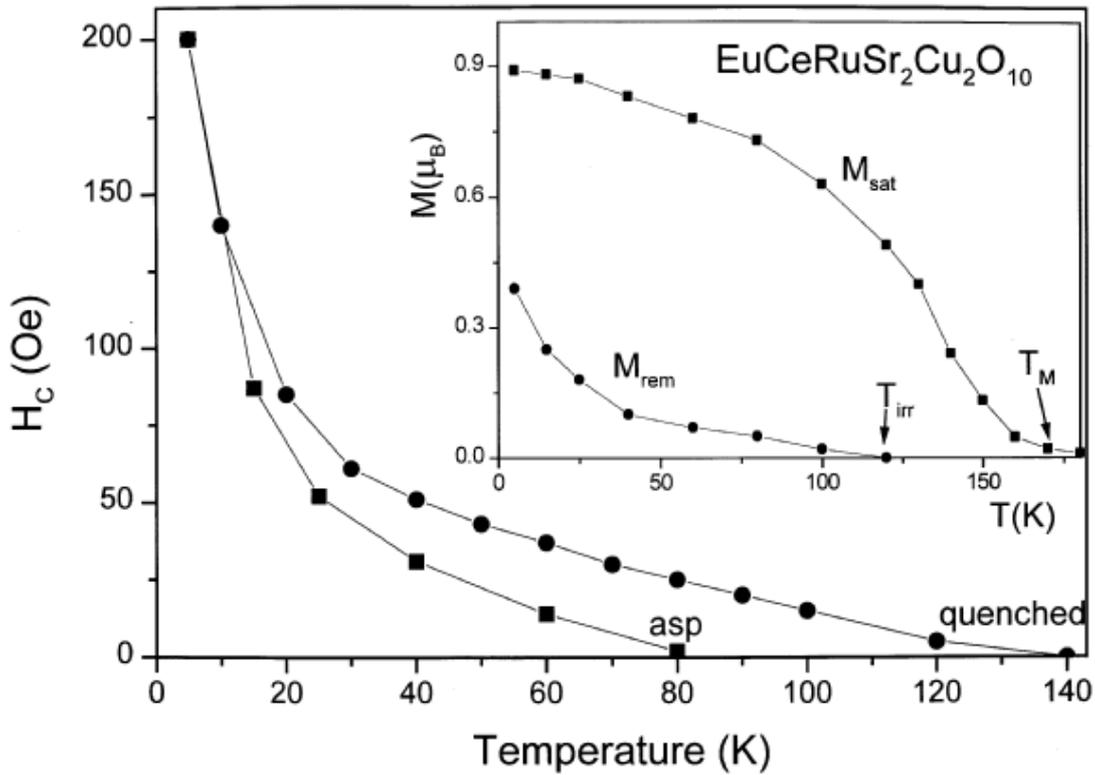

Fig. 5. The coercive field $H_C$ as function of temperature for asp and quenched $EuCeRuSr_2Cu_2O_{10}$ samples. The inset shows the temperature dependence of the saturation and the remanent moments.

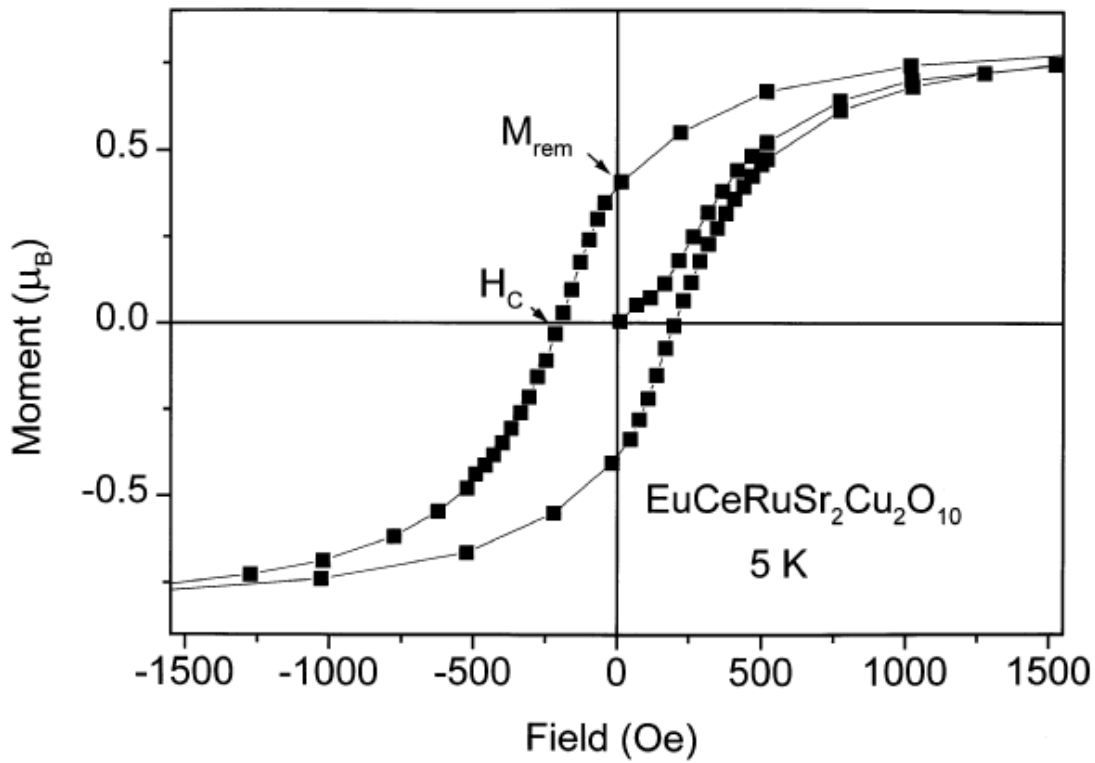

Fig. 6. The hysteresis low field loop at 5 K.

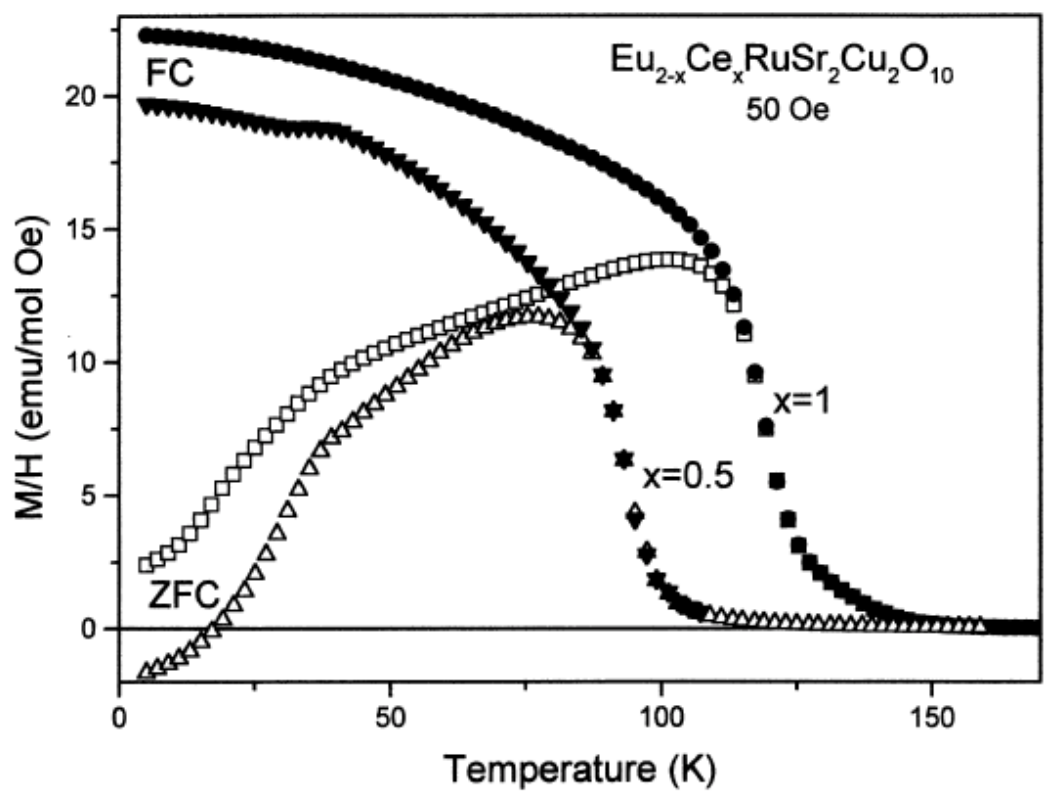

Fig. 7. ZFC and FC susceptibility curves for asp x=1 and x=0.5 samples, measured at 50 Oe.

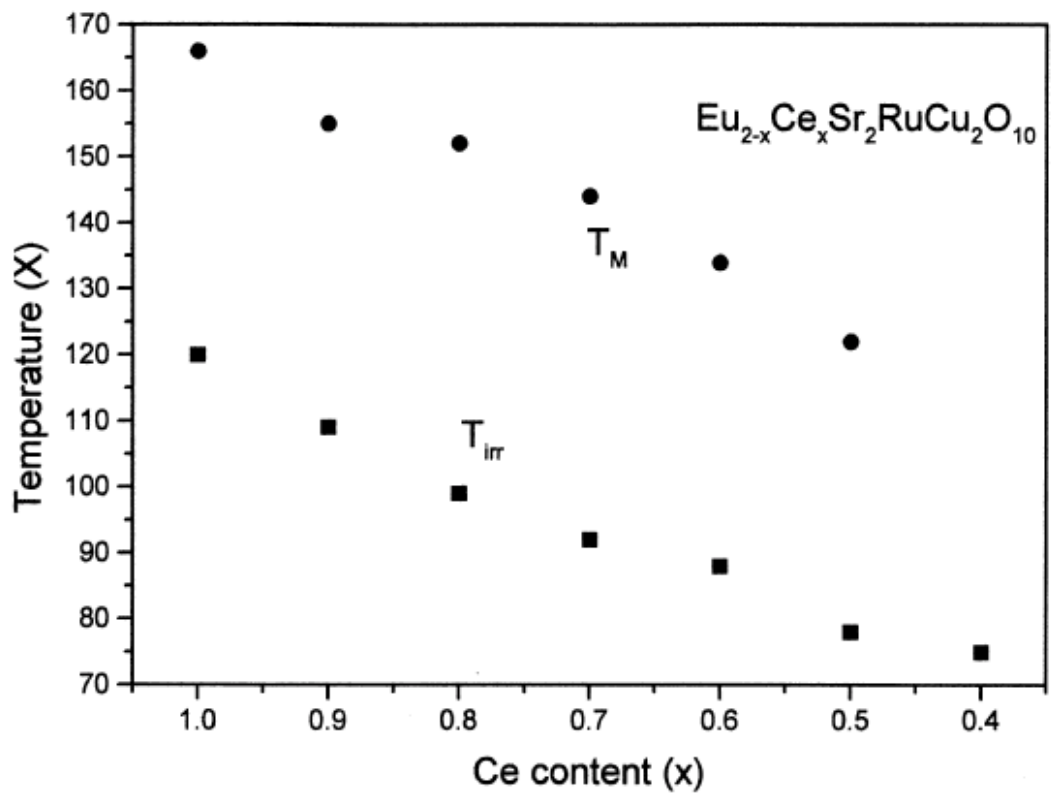

Fig. 8. $T_M$ and $T_{irr}$ as function of Ce in $Eu_{2-x}Ce_xRuSr_2Cu_2O_{10}$